\documentclass[prd,twocolumn,letterpaper,superscriptaddress]{revtex4}
\include{epsf}
\usepackage{graphicx}
\usepackage{pslatex}
\begin{document}

\title
{Accelerator Measurements of the Askaryan effect in Rock Salt:\\
A Roadmap Toward Teraton Underground Neutrino Detectors }

\author{P. W. Gorham}
\affiliation{Dept. of Physics \& Astronomy, Univ. of Hawaii at Manoa, 2505 Correa Rd.
Honolulu, HI, 96822}
\author{D. Saltzberg}
\affiliation{Dept. of Physics \& Astronomy, Univ. of Calif. at Los Angeles, Los Angeles, CA}
\author{R. C. Field}
\affiliation{Stanford Linear Accelerator Center, Stanford University, Menlo Park California}
\author{E. Guillian}
\affiliation{Dept. of Physics \& Astronomy, Univ. of Hawaii at Manoa, 2505 Correa Rd.
Honolulu, HI, 96822}
\author{R. Milincic}
\affiliation{Dept. of Physics \& Astronomy, Univ. of Hawaii at Manoa, 2505 Correa Rd.
Honolulu, HI, 96822}
\author{D. Walz}
\affiliation{Stanford Linear Accelerator Center, Stanford University, Menlo Park California}
\author{D. Williams}
\affiliation{Dept. of Physics \& Astronomy, Univ. of Calif. at Los Angeles, Los Angeles, CA}

\begin{abstract}
We report on further SLAC 
measurements of the Askaryan effect: coherent radio emission
from charge asymmetry in electromagnetic cascades. We used synthetic rock salt as the
dielectric medium, with cascades produced by GeV bremsstrahlung photons at the
Final Focus Test Beam. We extend our prior discovery measurements
to a wider range of parameter space and explore the effect in a dielectric medium of great
potential interest to large scale ultra-high energy neutrino detectors: rock salt (halite),
which occurs naturally in high purity formations containing in many cases hundreds of
km$^3$ of water-equivalent mass.
We observed strong coherent pulsed radio emission over a frequency band from 0.2-15 GHz. 
A grid of embedded dual-polarization antennas
was used to confirm the high degree of linear polarization and track the change of
direction of the electric-field vector with azimuth around the shower. Coherence was
observed over 4 orders of magnitude of shower energy. The frequency dependence of the
radiation was tested over two orders of magnitude of UHF and microwave frequencies.
We have also made the first observations of coherent transition radiation from the
Askaryan charge excess, and the result agrees well with theoretical predictions.
Based on these results we have performed a detailed and conservative 
simulation of a realistic GZK neutrino telescope array within a salt-dome, 
and we find it capable of detecting 10 or more contained events per year from even the
most conservative GZK neutrino models. 
\end{abstract}

\maketitle

\section{Introduction}
It is now widely understood that the universe becomes largely opaque to photons above
about 100 TeV due to pair production on the cosmic infrared 
and microwave background~\cite{GRpair}.
Stable charged baryons or leptons do not become magnetically rigid enough to propagate
over intergalactic distances until their energies are so high that they
also suffer significant losses through interactions with the cosmic
microwave background , through the process first noted by
Greisen, and Zatsepin \& Kuzmin (GZK) in the early 1960's~\cite{GZK,GZK1}.

Extragalactic astronomy in the $10^{15-21}$eV energy range at $\geq~0.1$~Gpc scales
must therefore utilize other messengers, and neutrinos are the
most likely contender. Moreover, the existence of sources that can produce single
particles with energies approaching 1~Zeta-electron-volt ($10^{21}$eV= 1 ZeV = 60 Joules)
provides a compelling driver toward development of 
ultra-high energy neutrino detectors, which can shed light not only on
acceleration processes at the sources that produce the primary cosmic rays, but on
the acute problem of their propagation and attenuation in the intergalactic medium
through the GZK process. 
This process itself yields neutrinos as secondaries of the $p\gamma_{2.7K}$
interactions, a process first noted by Berezinsky \& Zatsepin~\cite{BZ70}, 
and the resulting GZK neutrino spectrum~\cite{Engel01}, peaking at EeV energies,
distinctly reflects the cosmic ray source spectrum and distribution and provides 
unique diagnostics on the evolutionary history of the cosmic ray accelerators.
Because the GZK neutrino flux arises from the aggregate intensity of
all cosmic ray sources at all epochs, it is also potentially the strongest source
of EeV neutrinos integrated over the entire sky, and may in fact act as a ``standard
candle'' for EeV neutrinos.  Also, given the $\sim 100$ TeV center-of-momentum energy
of GZK neutrino interactions on nucleons at Earth, the possibility of high-energy physics
applications cannot be ignored. The combination of wide acceptance for the flux models with
a strong science motivation has made the detection of GZK neutrinos the focus of significant
attention for all current high energy neutrino detectors.

This goal compels us to consider detectors with target masses approaching $10^{42}$
nucleons, a Teraton of mass, or 
1000 km$^3$ water equivalent at EeV energies~\cite{SeckelRadhep}. Such a target mass
must be physically accessible, and must be able to transmit information about embedded
neutrino cascades within its enormous volume to a suitable and cost-effective
detector array. A promising approach toward this
daunting task is to utilize the Askaryan effect~\cite{Ask62,Ask65}, 
a process which leads to strong, coherent
radio pulses from such cascades. Several known dielectric media, with ice and
rock salt the most promising at present, appear to have the necessary
characteristics. Antarctic ice has been the focus of studies for this goal for
several years now via the Radio Ice Cerenkov Experiment (RICE)~\cite{RICE}, 
and the Antarctic Impulsive Transient Antenna (ANITA) long-duration balloon 
project will seek to exploit the effect in a novel approach~\cite{ANITA}. 
A recent limit at extremely high neutrino energies has been reported via FORTE satellite 
observations of the Greenland ice sheet~\cite{FORTE03}.
There is also 
another class of experiments searching for \v Cerenkov emission from UHE 
neutrinos via neutrino interactions with the lunar 
regolith ~\cite{Zhe88,Dag89,H96,GLUE}.  

Rock salt, first suggested as a possible target medium by Askaryan~\cite{Ask62},
has also been the focus of recent efforts, and it appears to show
equal promise with ice in regard to potential for detector development, 
along with the likely advantage of greater accessibility~\cite{Chiba04,NIMsalt}. 
We report here on experiments to further
understand aspects of the Askaryan process with a particular focus on 
rock salt as the medium for the cascades and radio pulse production. Our results
now extend measurements of the effect over the entire range of cascade
energies and radio frequency range of interest
for GZK neutrino detection. We have also explored in greater detail the
shower calorimetry aspects of radio \v Cerenkov measurements, and have made
the first measurements of the vector change of polarization at
different locations around the shower axis. This latter property in particular
is unique to radio \v Cerenkov detection, and the combination of all these
results gives us confidence that the road to Teraton neutrino detectors
is still wide open along this approach.

\section{Experimental setup}

\begin{figure} 
\epsfxsize=3.3in
\epsfbox{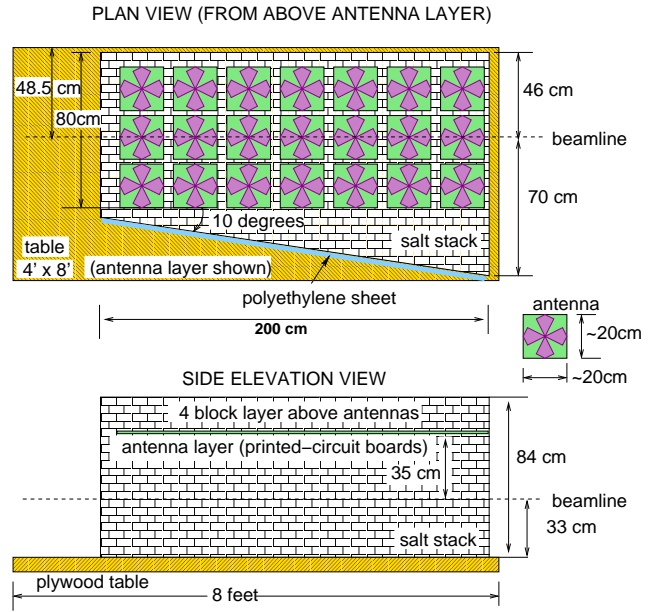}
\vspace{10pt}
\caption{Geometry of the salt-block shower target used in the experiment.}
\label{geom}
\end{figure}

The experiment (SLAC T460) was performed at the Final Focus Test Beam (FFTB) facility
at SLAC in June 2002. As in our first SLAC measurement~\cite{Sal01}, 
we used gamma-rays from a series of
Aluminum radiators of different thicknesses to provide secondary bremsstrahlung
gamma-rays for shower production in our rock salt target. The geometry
of the target is shown in Fig.~\ref{geom}. It was built of 1.8 kg salt bricks
obtained from Morton Salt Inc., with a total mass of about 4 metric tons. 
The bricks are slightly trapezoidal in all of their cross sections to
accommodate the manufacturer's mold release, 
but have average dimensions of $15 \times 10 \times 6$~cm.
Such bricks are food-grade, and specified to be at least 99.5\% pure
sodium chloride. They are manufactured by compression molding of crystalline salt
under $\sim 450$~tons per in$^2$ pressure. Their density was measured to be 2.08 g cm$^{-3}$,
which is about 3\% less than the density of pure halite mineral. 

Measurements
of the water content of small samples of a typical brick found only trace 
amounts of volatiles (including water), below 0.17\%. 
Laboratory measurements of sodium chloride crystals
at comparable purities give upper limits of $\leq 10^{-4}$
on the loss tangent $\tan{\delta}$, which is related to the 
absorption coefficient $\alpha$, for small values of $\tan{\delta}$, by
\begin{displaymath}
\alpha \simeq \frac{2\pi\nu}{c}~ \sqrt{\epsilon} ~\tan\delta~~~{\rm (m^{-1})}
\end{displaymath}
for frequency $\nu$, real part $\epsilon$ of the dielectric permittivity,
and speed of light $c$.
The implied upper limits on the
attenuation length $L_{\alpha} = \alpha^{-1}$ 
are hundreds of meters~\cite{vonHippel} in the UHF and
cm wave regime, and our tests through
of order 0.3-0.4~m of the salt bricks found no measurable attenuation. 
In the results reported here we assume negligible bulk absorption of the radio
emission. Paper labels attached to the the top of the bricks were 
found to be difficult
to remove, and after confirming that they were non-conductive and 
had no effect on the  radio transmission, they were left alone.

To minimize the effects of the slight gaps left when stacking the salt
blocks, pure (food grade) un-iodized table salt was used to fill in the 
interstices. A 2.5~cm thick polyethylene sheet was also mounted along the
Cherenkov radiator surface to provide a smooth refracting surface for the
external antennas and some improvement in index matching because of the
intermediate RF index of refraction of polyethylene.

An array of 7 by 3 (21 total) printed-circuit board (PCB) broadband 
dual-linear-polarization bowtie antennas was embedded in
the upper portion of the target, on a rectangular grid with spacings of
22 cm along the shower axis and 21 cm transverse to it, with one set of
antennas arranged with one of their polarization axes aligned directly
above the beamline. These are shown schematically in Fig.~\ref{geom}.
The antennas had a frequency response of approximately
0.2-2 GHz, and about -12 dB of cross-polarization rejection. 

\begin{figure} 
\epsfxsize=3.3in
\epsfbox{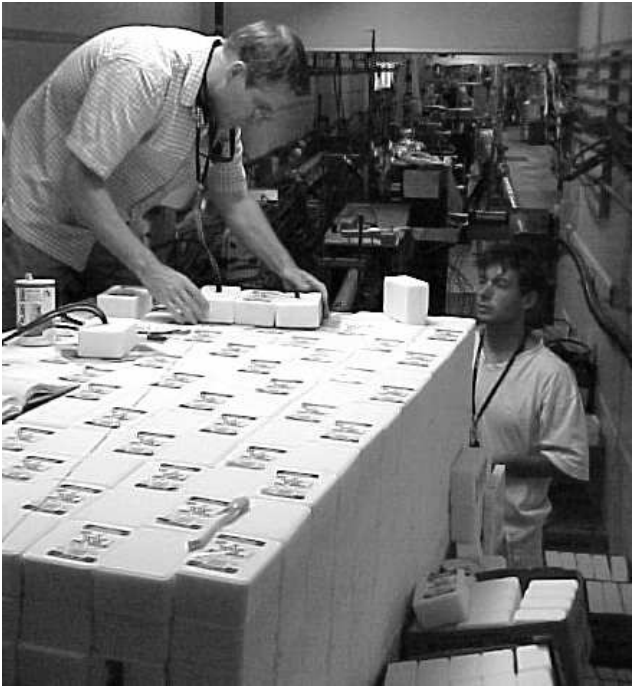}
\vspace{10pt}
\caption{View of the salt stack construction at the antenna layer. Antennas
were mounted with their leads passing through drilled salt blocks as
seen in several examples on the top of the stack.}
\label{stack}
\end{figure}

In addition
to the embedded array, we also used two external antennas, a C/X-band
standard gain horn usable over the frequency range from
5 to 9 GHz, at a fixed location on the side of the target; and a 1-18 GHz
log-periodic dipole array (LPDA) antenna with a single linear polarization,
which was also located to the side of the target but could be repositioned.
One side of the salt stack was arranged with a $10^{\circ}$ angle with
respect to the beamline to facilitate transmission of the \v Cerenkov radiation,
which would have otherwise been totally internally reflected from a 
parallel face. The partially constructed salt stack is shown in Fig.~\ref{stack}
at the antenna layer, with several of the antennas evident, assembled with their
leads projecting through a drilled salt brick.

All antenna signals were digitized with Tektronix digital oscilloscopes 
(a TDS694C 3~GHz bandwidth, 10~Gsamples/s; and a CSA8000 sampling scope,
20~GHz, up to 50 Gsamples/s) using an
ultra-stable microwave transition-radiation-based trigger from an upstream location.
Runs were also taken with the bremsstrahlung radiators in and out of the beam
at numerous times during the data collection to test for the presence of
accelerator backgrounds, which were found to be negligible.
The total shower energy was varied by changing both the electron beam
intensity and the thickness of the bremsstrahlung radiators. These radiators
varied from  0.06\% to 1.5\% of a radiation length
and thus extracted 
the same fraction of the electron beam energy in bremsstrahlung
photons. The electrons were then steered away from the salt target and dumped
into the FFTB beam dump approximately 20~m downstream. The  
bremsstrahlung photon spectrum from the radiators used is very broad
with a mean energy of several GeV.

SLAC beam current
monitors allowed us to reduce beam intensity to about $10^9$ electrons per
bunch using conventional methods. Below this level, closed loop
control (eg. fine-tuning) 
of the beam direction was not possible. However, by employing 
our high signal-to-noise
coherent microwave transition radiation trigger, we were able to further
monitor the beam intensity to levels an additional 2 orders of magnitude
below this. With the combination of radiators available to us, the
dynamic range of the showers obtained was 
thus about 4 orders of magnitude in shower energy. 
The excellent stability of the FFTB beam assured that no
retuning was necessary during our measurements, which were repeatable once
the beam intensity was set back to the higher levels again.

The showers produced in such experiments consist of the superposition of a
large number of GeV electromagnetic cascades, with energies that sum up to EeV
levels, but whose development is determined by the convolution of a GeV
cascade and the exponential first-interaction distribution of the input
photons. As such it cannot simulate the leading interactions of a cascade
initiated by a single high energy particle, either in the longitudinal
shower distributions of particle number, or in the exact particle species
content. However, these differences are of little consequence for radio
production, since the Askaryan emission is dominated by the electron
excess at energies below $\sim 100$~MeV, near the region of shower
maximum. The charge excess itself is quite similar in either case, and
the main quantitative differences come in only as the logarithm of
the ratio of the lead particle energy to the critical energy, which
produces a more extended shower maximum region in the case of a shower
initiated by a high energy particle, and gives more total track-length
for Cherenkov production. These differences are easily
quantified, and in practice the GeV composite showers used here can
be used to rigorously test models for the Askaryan radio emission.

\section{Results}

To provide a baseline for evaluation of the results, we simulated the 
electromagnetic showers using the Electron Gamma Shower 4 (EGS4) Monte Carlo
code~\cite{EGS4}. EGS4 cannot directly simulate the radio emission
from the showers, but can assess both longitudinal and lateral development
of the total charged particle and photon content of the shower, as
well as the Askaryan charge excess. 

\begin{figure} 
\epsfxsize=3.5in
\epsfbox{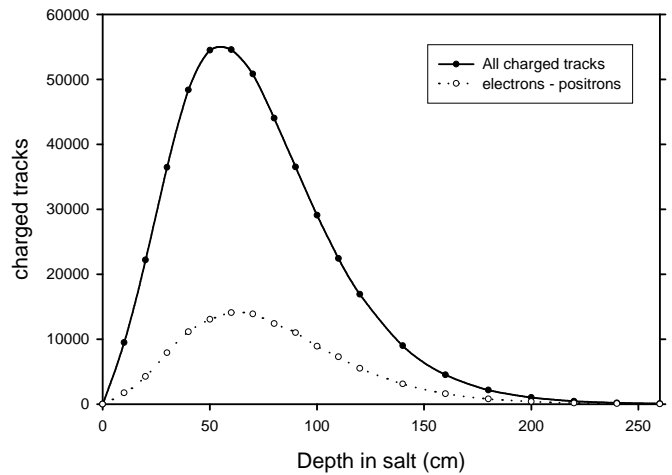}
\vspace{10pt}
\caption{Longitudinal distribution of the bremsstrahlung shower, for both
total charge and charge excess as simulated by EGS4 for rock salt.}
\label{longdist1}
\end{figure}

\begin{figure} 
\epsfxsize=3.5in
\epsfbox{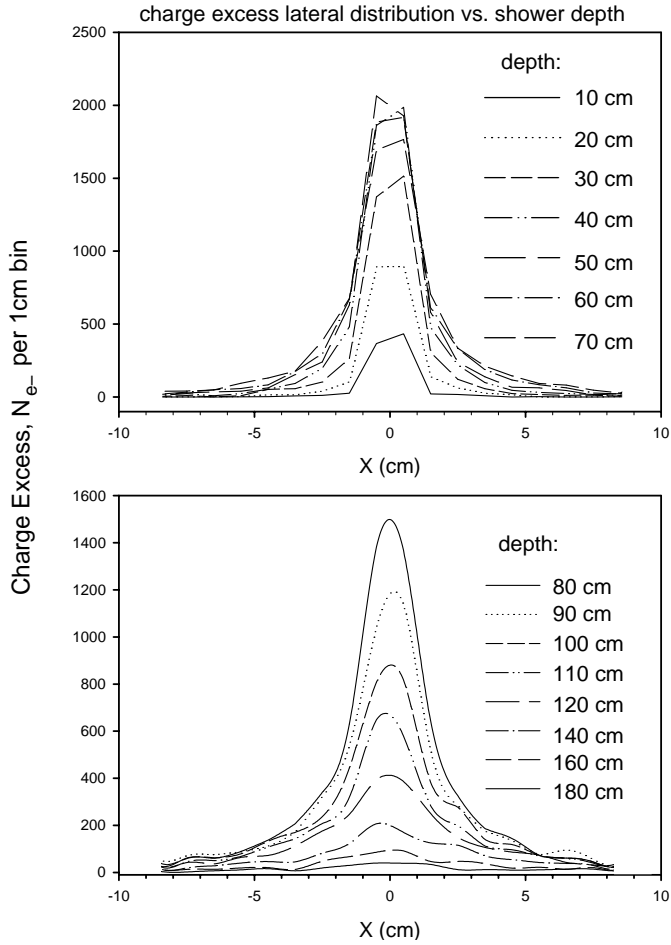}
\vspace{10pt}
\caption{Lateral distributions at various depths. Top: depths less than
70~cm. Bottom: depths greater than 70~cm.}
\label{latdist}
\end{figure}

Figure~\ref{longdist1} shows the longitudinal
shower development for a salt target, including the charge excess.
Fig.~\ref{latdist} shows a series of profiles of the shower lateral distribution
of the excess charge 
at different depths in cm, indicated on the plot. In each case the simulation is for 
1000 28.5 GeV electrons initially, led through a 10\% bremsstrahlung radiator to
fully account for the secondary photon spectrum. No changes are
required to the EGS4 code to develop the $\sim 25\%$ negative charge excess seen;
this again demonstrates that the Askaryan effect is generic to
electromagnetic showers.

\begin{figure} 
\epsfxsize=3.3in
\epsfbox{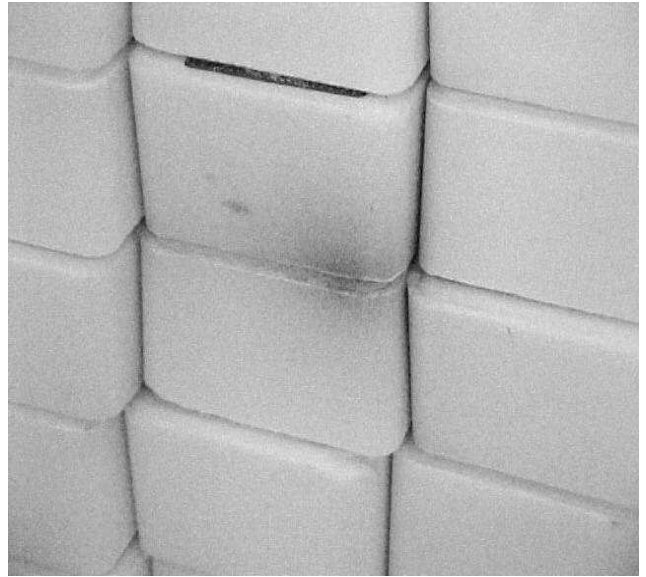}
\vspace{10pt}
\caption{The exit point of the beam in the downstream wall of the salt
target, outlined by discoloration of the salt produced by beam and
showing in cross section at the exit surface. Even at this location
of order 15 radiation lengths along the shower, the shower
core is still only several cm across (the salt blocks are 6 by 10 cm
here in cross section).}
\label{beamexit}
\end{figure}

The simulations of the shower transverse distribution were qualitatively
confirmed by the appearance of a deposit of discolored material  along
a volume of salt within the stack
during the experiment, the evidence of which is shown in Fig.~\ref{beamexit}.
The size and shape of this cross section of the contaminated region is 
an indication of the size of the shower core itself, and it is still
of order several cm in diameter even at a depth of $\sim 15$ radiation lengths.

\subsection{Coherence \& Absolute Field Strength.}

In reference~\cite{Sal01}, the coherence of the radiation was observed over a fairly limited range
of shower energies, from about $3 \times 10^{17}$~eV to $10^{19}$~eV composite
energy. In our current measurements, we have extended the coherence measurements to
a much larger range. These measurements were made using averaged results from several of
the bowtie antennas around the region of shower maximum. Typical runs at a given total
charge per bunch (which determines the composite energy per bunch) were averages of
1000 beam shots, and the beam current was separately monitored for stability. 
Typically we also recorded 100-1000 shot-to-shot pulse profiles to monitor the variance.
Once corrected for minor variations in beam current, these were stable to the level of
a few percent.

\begin{figure} 
\epsfxsize=3.5in
\epsfbox{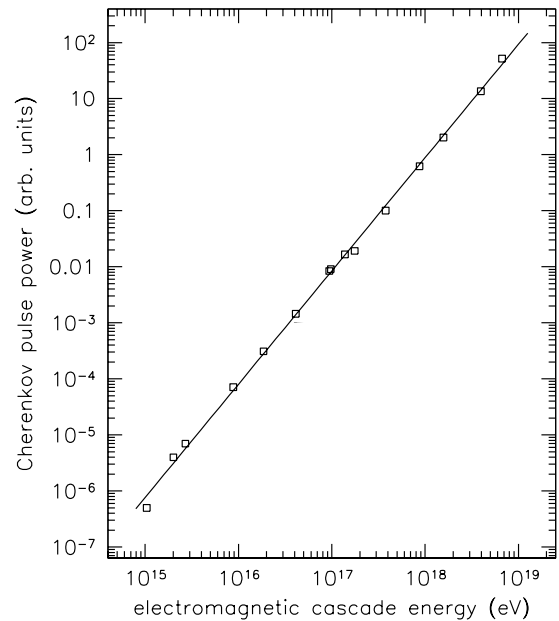}
\vspace{10pt}
\caption{Observed coherence of the 0.3-1.5GHz radiation as a 
function of total beam energy per pulse. The curve shows a quadratic
relation for power as a function of shower energy.}
\label{coh02}
\end{figure}

Fig.~\ref{coh02} shows the results of the relative radio-frequency power as a function of
shower energy, now covering nearly 4 orders of magnitude in energy and 8 orders of magnitude
in RF power. No departures from the expected quadratic dependence of pulse power with
shower energy are seen. This result demonstrates the remarkable dynamic range possible
for detection of coherent RF \v Cerenkov radiation.



Determination of the absolute measured field strength of the showers requires that the
antenna effective area, spectral response, coupling efficiency, and angular response 
all be included in converting the measured voltages to field strengths in the 
detection medium. To properly account for the angular response, one must first
determine if the conditions for the reception are near-field (Fresnel zone) or
far-field (Fraunhofer Zone) for both the antenna and radiation source. The conditions
for far-field response are approximately
\begin{equation}
f \leq {Rc \over L^2\cos^2 \theta}
\end{equation}
where $R$ is the distance between source and antenna, $L$ the largest antenna dimension,
$\theta$ angle of incidence with respect to the normal (broadside) of the antenna, and
$f$ is the highest frequency for which far-field conditions are met.
For the antenna as a receiving component, with $R = 38$~cm, $L=20$~cm, and
$\theta = 25^{\circ}$ at the \v Cerenkov angle
the antenna far field conditions are satisfied for 
frequencies below 2~GHz, the highest usable frequency
for the bowtie antennas.

\begin{figure} 
\epsfxsize=3.5in
\epsfbox{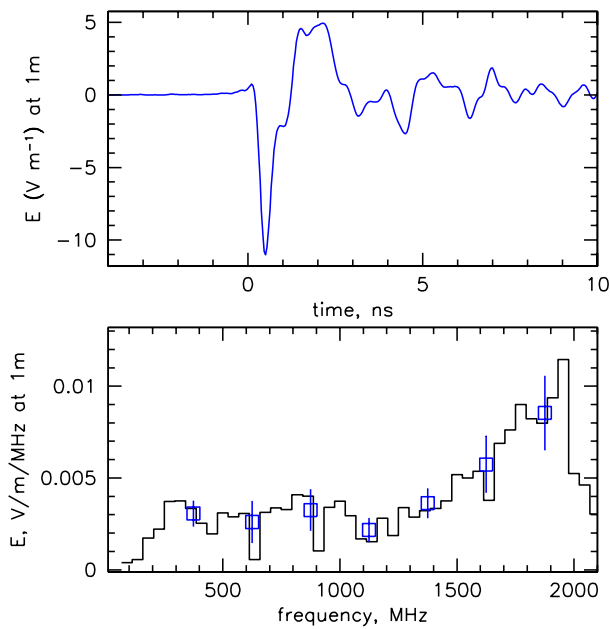}
\vspace{10pt}
\caption{Top: Summed-in-phase pulse profile from all bowtie antennas. 
Bottom: Absolute power
spectrum of the bowtie data, corrected for antenna angular response.}
\label{allpower1}
\end{figure}

While the shower is in the far-field of the antenna, the converse
is not the case, since the radiation develops over a region
comparable in size to the entire shower with a 
length of about 80~cm for full-width-at-half-maximum
near the peak of cascade particle development.
In practice this means that while the far-field antenna
beam patterns may be used to determine the bowtie amplitude response with angle,
they receive radiation only from approximately that portion of the shower that
is projected onto the antenna effective area at the \v Cerenkov angle (about $66^{\circ}$).
This behavior was observed in ref.~\cite{Sal01}, 
and provides the possibility of tracing out the
longitudinal charge evolution of the shower by 
measuring the change in radiation amplitude
along the shower axis, as was done in ref.~\cite{Sal01} 
(see also Fig.~\ref{stokesv} below). In determining the absolute field
strength one must therefore account for the fact that the shower is
effectively resolved into these individual sections.

Figure~\ref{allpower1}(top) shows a plot of the summed field strengths from all of the
bowtie antennas, corrected for known cable attenuation but not for other
frequency dependent effects. The pulse shape is typical of the impulse response for
bowtie antennas.  In the bottom portion of the figure, a Fourier amplitude spectrum of
the field strength is shown, including corrections for the gain of the bowtie antennas.
The solid histogram shows the power spectrum of the pulse, and the points are
averages over 250~MHz sections of the spectrum, with the error bars based on the
spectral variance within each section. It is evident that there are some near-nulls in
the bowtie response, but the overall trend is for a rise at the higher 
frequencies, consistent with \v Cerenkov radiation.

\begin{figure} %
\epsfxsize=3.3in
\epsfbox{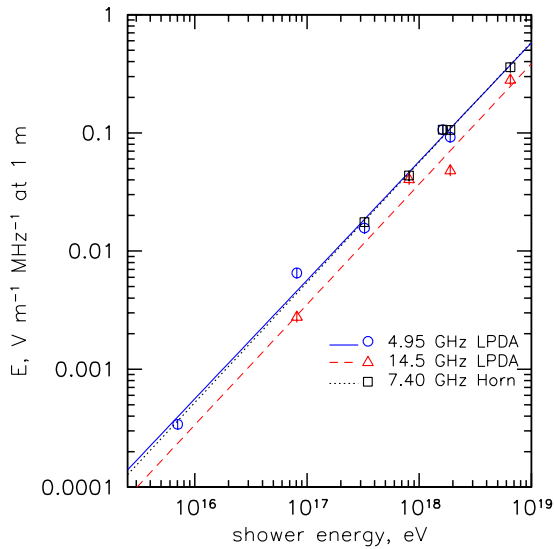}
\vspace{10pt}
\caption{ Measured coherence of electric field strength at 4.95, 14.5 with 
LPDA and 7.4 GHz with horn antennas respectively, with least-squares fit curve.} 
\label{coh_freq02}
\end{figure}

We also measured the coherence at higher frequencies (2.2-15.0 GHz) 
in several bandpasses, using the LPDA
and the C/X-band horn antennas, both of which had known effective areas.
Fig.~\ref{coh_freq02} shows the measured electric field strength vs. total shower 
energy at different frequencies. Measurements were done with the LPDA on 4.95 and 
14.5 GHz, respectively and with the horn antenna at 7.4 GHz. Points represent 
measured values and lines are least-square fit curves 
$| \mathbf{E} |=A {E_{sh}}^{\alpha}$, where $\mathbf{E}$ is the electric
field and $E_{sh}$ the shower energy. 
The fit for the exponent ${\alpha}$ gives: ${\alpha_{4.95}} = 1.00 \pm 0.04$ at 4.95 GHz, 
${\alpha_{14.5}} = 1.02 \pm 0.11$ at 14.5 GHz, ${\alpha_{7.4}} = 0.99 \pm 0.05$ 
at 7.4 GHz. On all three frequencies, the field strength is in agreement with full 
coherence of the radiation.

\begin{figure} %
\epsfxsize=3.3in
\epsfbox{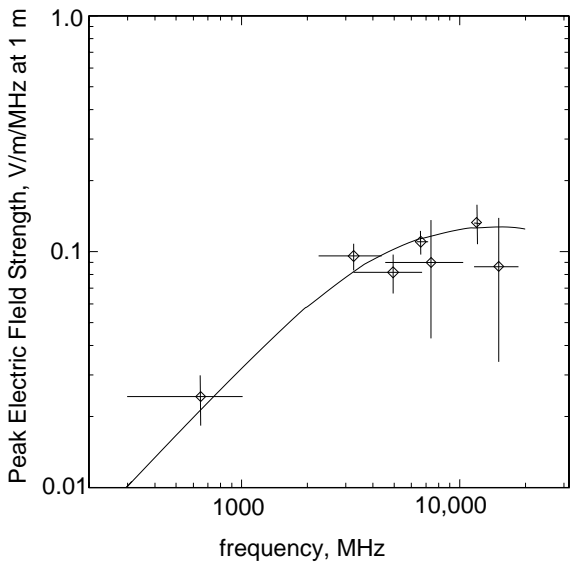}
\vspace{10pt}
\caption{ Spectral dependence of the measured electric field strength 
for shower energy $1.9\times 10^{18} eV$. The solid curve is a parametrized 
Monte Carlo simulation scaled for synthetic rock salt. Vertical bars are
estimated errors, largely due to systematics in the absolute RF calibration;
horizontal bars indicate the bandwidth used.}
\label{spectrum02_1}
\end{figure}

Fig.~\ref{spectrum02_1} shows the absolute field strength measured 
in several frequency bands 
from 0.3-15.0 GHz. The plotted curve is based on the parametrization 
given in~\cite{Alv98,Alv97}, 
scaling from ice to synthetic rock salt 
(for details see Appendix B). 
The horizontal bars are not standard errors but show antenna 
or filter bandwidth, and 
vertical bars indicate combined statistical and systematic uncertainties. 
The measured spectrum of the pulse electric field is 
in good agreement with the prediction.     

\subsection{Polarization and charge excess tracking}

The crossed bowtie antennas employed in our measurements provided simultaneous measurements
of the induced voltage in orthogonal linear polarizations, but because of their geometry,
there is inherent leakage of a co-polarized signal into the cross-polarized receiver. With
a measured 12 dB of cross-polarized rejection, up to 15\% of the co-polarized amplitude
(eg. voltage) can appear in the cross-polarized channel. To make accurate estimates of 
the angle of the projected plane of polarization of the radiation, the polarization leakage must
be accounted for. We use a simple model for this, where the co- and cross-polarized voltages
are related by:
\begin{eqnarray}
V_{0}  ~=~ {\bf E}\cdot [ {\bf h_{0}}  + \alpha{\bf h_{90}} ] \\
V_{90} ~=~ {\bf E}\cdot [ {\bf h_{90}} + \alpha{\bf h_{0}}  ] 
\end{eqnarray}
where ${\bf E}$ is the electric field vector of the radiation and 
${\bf h_{0,90}}$ are the
vector effective heights (which are complex in general) of the co- 
and cross-polarized
antennas. Since each of the bowtie antennas are identical in both the co- and cross-polarized
directions, and since the induced voltage has a simple sinusoidal 
dependence on
the projected plane of polarization angle $\Psi$ 
(assuming no net circular polarization), we can write
\begin{eqnarray}
V_{0}  ~=~  Eh (\cos\Psi + \alpha \sin\Psi ) \\
V_{90} ~=~  Eh (\sin\Psi + \alpha \cos\Psi ) 
\end{eqnarray}
where $h = \frac{1}{2}(|h_{0}| + |h_{90}|)$ 
has the same magnitude for both the co- and cross-polarized antennas.

In our case we choose the co-polarized direction to be aligned with the beam
axis, since the radiation is naturally expected to be polarized along this axis. 
If we assume that $\Psi$ is approximately aligned with the beam axis (in practice
this was done to a precision of about $1-2^{\circ}$, then for an antenna along the
beam axis, we find to a good approximation that $\alpha = (V_{90}/V_{0})|_{on-axis}$.
$\Psi$ is then determined by
\begin{equation}
\Psi = -\tan^{-1} \left ( {\alpha - V_{90}/V_{0} \over \alpha V_{90}/V_0 - 1 } \right )~.
\end{equation}

\begin{figure} 
\epsfxsize=3.5in
\epsfbox{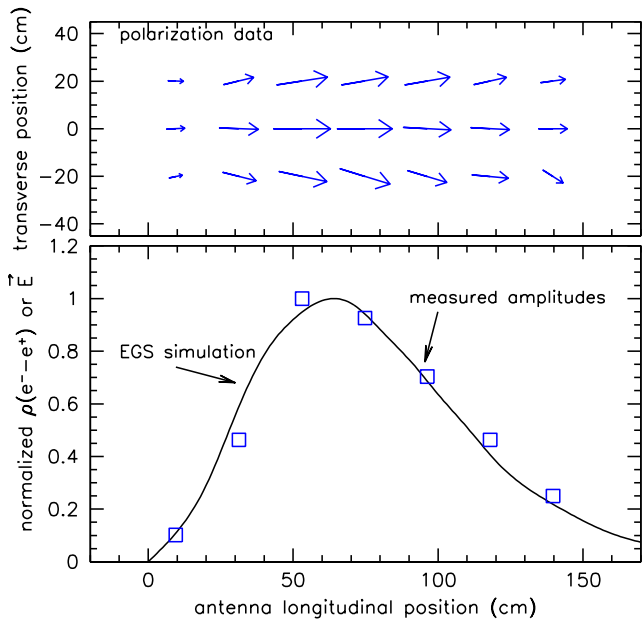}
\vspace{10pt}
\caption{Top: Measured polarization vector field at the bowtie antenna locations
in a plane 34 cm above the shower axis. The antenna feed center locations in that 
plane give the base of each vector, which is scaled in length 
as the square root of amplitude for
clarity. The angles of the vectors match the observed polarization angles.
Bottom: The observed amplitude of the received RF pulses along the center line of the
antenna array. Also plotted is a curve of an EGS simulation of these showers; both curves are
normalized to unit peak amplitude. }
\label{stokesv}
\end{figure}

\begin{figure} 
\epsfxsize=3.5in
\epsfbox{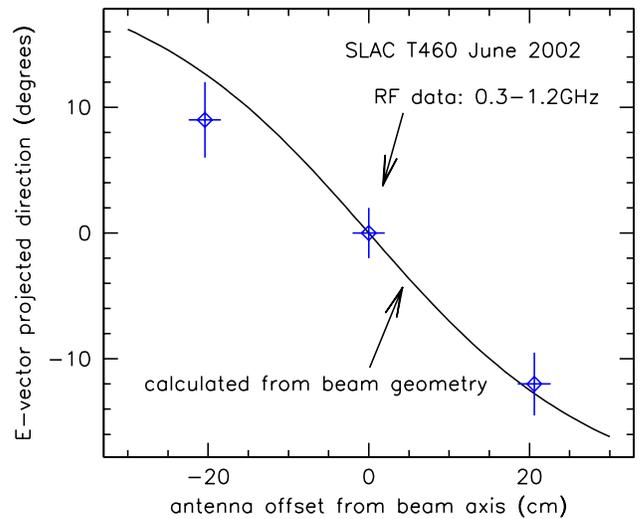}
\vspace{10pt}
\caption{Plane of polarization vs. transverse distance of the antennas, which were all about
35 cm above the axis of the beam.}
\label{pangle}
\end{figure}

Fig.~\ref{stokesv} shows results of the polarization and relative amplitude measurements
for all 21 of the bowtie antennas. The top pane plots these as vectors with scaled 
lengths and directions corresponding to the square root of the amplitude (used for
clarity for the lower amplitudes) and the projected plane of the polarization $\Psi$.
In the lower pane of the figure, we also show the relative amplitude of the antennas along
the center axis as a function of the longitudinal distance along the shower (corrected for
the \v Cerenkov angle projection). We also plot the EGS4 simulation of the shape of the charge excess
along the shower (about 27\% of the total charge).
There is excellent agreement between the measured shape of the amplitude
response and the normalized charge excess predictions from EGS4.

In Figure~\ref{pangle} we show similar data for the three antennas at shower maximum
at a depth of 50~cm, now plotting
the plane of polarization as a function of the transverse position of the antenna with
respect to the centerline of the beam. All of the antennas lay in a plane about 35 cm above the
beamline. The solid curve shows the expected change in the angle of polarization for
the three antennas, and the agreement is good.

\subsection{Transition radiation}
Transition radiation (TR) and \v Cerenkov radiation are closely related; this is particularly
true in the radio regime, where both forms of radiation may coexist in partially or
fully coherent forms, and interference between them may result. In earlier work ~\cite{Gor00}
we demonstrated the existence of free-space coherent microwave TR from relativistic
electron bunches exiting an accelerator beam pipe, and confirmed the predicted
angular dependence and polarization of the radiation. However, there was a serious
discrepancy in these earlier measurements with regard to the total emitted power,
which was observed to be significantly lower than predictions.

Transition radiation in the radio regime is of particular interest for detection of high energy
particle showers, since the Askaryan charge excess guarantees not only coherent \v Cerenkov
radiation, but also TR when there is any change in the dielectric medium on scales
comparable to the size of the shower. Showers which break through a solid surface can
produce forward TR in the direction of propagation, and showers originating in
one medium (air for example) which then intersect a solid surface will produce backward
TR which propagates opposite to the original shower direction. It is thus of
some importance to establish the emitted power from this process, to determine
under what circumstances it may be useful for high energy shower detection.

\begin{figure}
\epsfxsize=3.2in
\epsfbox{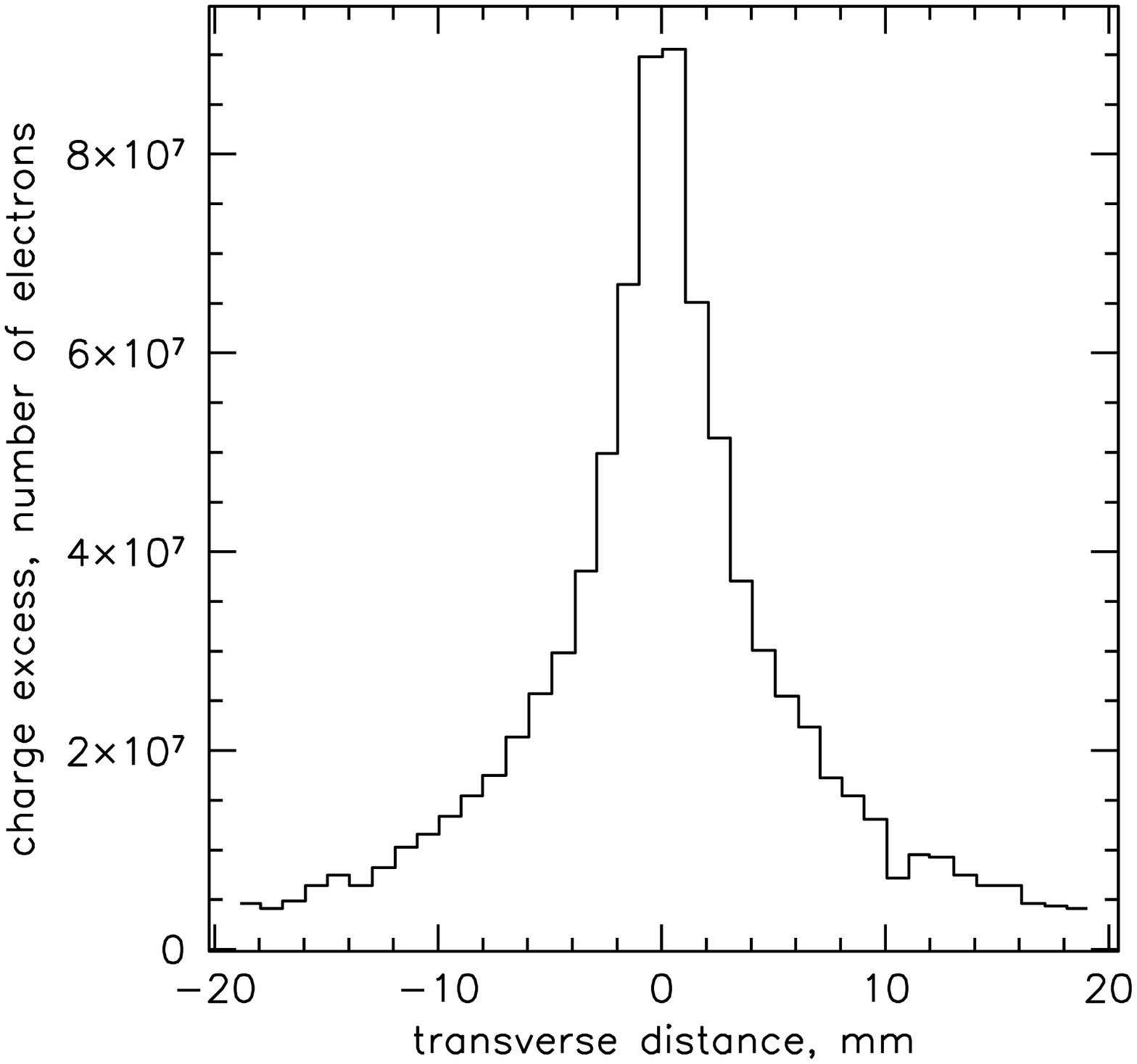}
\epsfxsize=3.2in
\epsfbox{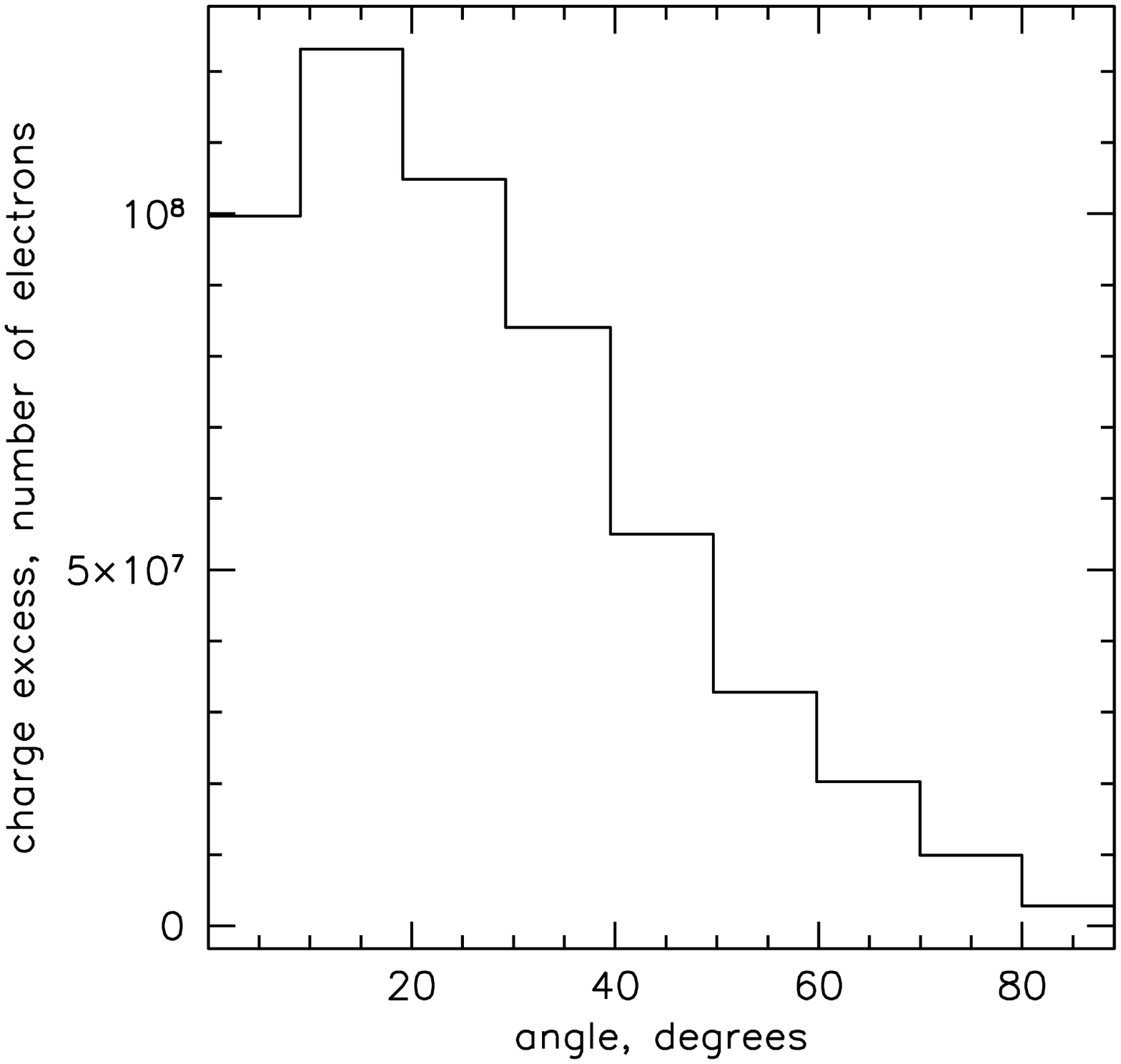}
\caption{Top: Transverse spread in 
the simulated electron shower emerging from the TR radiator,
grouped in 1~mm annular bins around the beam axis. 
Bottom: angular emittance spread in the same electron shower,
for the highest energy bin ($0.975 < \beta < 1$), in
annular bins of solid angle, with $10^{\circ}$ angular
width.\label{TR}}
\end{figure}

In a setup that was independent of the salt target, and done after the work with the
salt was complete, we used the photon beam to produce an electron-gamma shower 
in a 2 cm block of lead followed by a 1.3 cm block of aluminum. 
The excess electrons from the shower are thus expected to produce TR as they
exit the aluminum. Strong RF pulses were observed with an X-band horn, with
typical duration of 0.45 ns, consistent with a completely band-limited signal
($h_{eff} = 2.4$~cm, $\langle \lambda \rangle =4.2$~cm, $d\nu = 2$~GHz, 
effective area $A_{eff}$ = 18.87~cm$^2$). 
An EGS simulation
provided us with an estimate of the number of excess electrons, their energy, and
their angular and transverse distributions, which are shown in Fig.~\ref{TR}. In an appendix,
we provide details of the calculation of the expected TR power, which depends on
the geometric form factors of the exiting bunch.
The transverse and angular divergence parameters are estimated to be
$\sigma_T = 6$ mm and $\Omega = 30^\circ$, respectively; therefore $f_T = 0.996$
and $\chi = 0.08$. 
Most of the shower electrons are in the
highest energy bin ($0.975 < \beta < 1$), for simplicity we use only the electrons in
that bin for this calculation, setting the average $\beta = .9875$.
We assume the longitudinal form factor $f_L = 1$ which is
reasonable given the very tight longitudinal distribution of SLAC bunches. 

The results of the TR run are shown in Table~\ref{TRtable}.
\begin{table}[h]
\caption{TR measurements \& analysis.\label{TRtable}}
\begin{center}
\begin{tabular}{|c|c|c|c|c|c|c|c|}
\hline
\hline
$\theta$& D&$d\Omega$& $N_e$& $P_L$& $P_L$& $E$ & $E$ \\
& cm &  msr & & mW & mW & V/m &  V/m \\
 & & & & (calc.)& (meas.) & (calc.) & (meas.) \\
\hline
10$^{\circ}$& 135.5& $1.03$& $5.3\times 10^8$& 0.26 & 0.50 & 0.28 & 0.31\\
\hline
\hline
\end{tabular}
\end {center}
\end{table}

Here, the predicted power is about half the measured power, much closer than
in our previous result ~\cite{Gor00}. The predicted electric 
field is 90\% of the measured electric field.
We would expect it to be about 70\% of the measured electric field based on the
factor of 2 difference in the power, but the electric field estimate assumes
a perfectly triangular envelope, which is not accurate. Detailed studies indicate that
the peak and RMS voltages are related by $V_{RMS} = kV_p/2\sqrt{2}$, 
where $k \approx 1.4$~\cite{DawnThesis}.

\section{Discussion}
Our measurements have extended the validity of the theory underlying
the Askaryan effect to a second dielectric material, with 30\% higher
density compared to silica sand, and a 40\% higher dielectric constant.
In all cases the theory appears to scale as expected by the material
properties, and there is no reason to believe this scaling would
not apply for other dielectric materials as well. The measurements
of transition radiation from the charge excess are the first 
observations of this effect, and lends further strength to the Askaryan
hypothesis that a negative charge excess is responsible for the
coherent radio emission, for if this were not the case, there is no
obvious way to produce the measured transition radiation with the
strength observed.

This latter effect could be important for both cosmic ray shower and
neutrino shower detection, since there are many physical situations
where a shower encounters a discontinuity of some kind that can lead 
to transition radiation. For example, ultra-high energy air showers
encountering clouds will see a dielectric discontinuity; and
giant air showers which impact the earth or the surface of the ocean
will certainly produce strong backward transition radiation, 
particularly in the latter case where the Fresnel reflection coefficient
from the surface of the ocean is of order unity.

Our radio \v Cerenkov measurements
demonstrate conclusively that coherent \v Cerenkov radio pulses from 
showers in the rock salt can accurately reconstruct the profile of the shower
development as well as the total shower energy. They show that the process
is inherently quadratic in the rise of power vs. shower energy, and this
relationship is reliable over many decades of these parameters.
Combining these results with  {\it in situ} measurements of propagation
characteristics  at radio frequencies in salt domes~\cite{NIMsalt}
which indicate $\ge 250$~m attenuation length, we conclude that nothing
in principle prevents the development of embedded antenna arrays within large
salt structures such as salt domes for the characterization of ultra-high energy
neutrino fluxes.

\subsection{Application to detector modeling.}
To further understand how this technique might be applied, we have created
a Monte-Carlo simulation of a large-scale antenna array embedded within a 
generic salt dome,
where the halite with the highest purity is likely to reside.
We denote the array here as a Salt dome Shower Array (SalSA). 
The array is cubic with
1728 antenna nodes ($12 \times 12 \times 12$), with a grid spacing of 225 m,
of the same order as conservative {\em in situ} estimates~\cite{NIMsalt} 
of the field attenuation length in rock salt,
which we assume to be 250~m in this simulation. In fact 
recent laboratory measurements using a dielectric cavity~\cite{Chiba04}
on samples of rock salt from the Hockly salt dome favor $L_\alpha \simeq 900$~m 
at 200~MHz, and similar long attenuation lengths are also favored by
ground-penetrating radar data~\cite{NIMsalt}.

Each node consists of 12 closely spaced
antennas (0.75 m vertical offset), 6 of which are dipoles with vertical polarization,
and 6 of which are slotted-cylinder antennas which are sensitive to horizontal
polarization. The antennas are both assumed to have $\cos^2 \theta$
dipolar response functions
for simplicity here. The assumption is accurate for fat dipoles or bicones in the
vertical polarization~\cite{Kra88}; for our slot cylinders, the response in practice is 
somewhat more flattened 
than $\cos^2 \theta$ with a 1-2 dB front-to-back asymmetry, but we
anticipate further development of current designs will tend to
converge on a more uniform dipolar response.

Multi-antenna local nodes such as this are effective in enabling
a lower local-trigger threshold in the presence of Gaussian thermal noise,
and are very cost effective since the antenna elements themselves are
inexpensive. The antennas are based on
Numerical Electromagnetics (NEC2) models and actual tested prototypes, and
are optimized for frequencies centered at about 200~MHz, with bandwidth of about
100\%. Beam patterns for the mid-frequencies are assumed.

A system temperature of 450K is assumed, based on about 310K for
the salt and a receiver noise temperature of 140K, consistent with low-noise
amplifiers that are readily available commercially. The basic design for node
electronics centers on the use of a switched-capacitor array (SCA) transient
digitizer, which is only read out and sent to the
surface when the local node is triggered; a prototype of the basic design,
which uses Gigabit ethernet on one fiber per node to the surface, is
described by G. Varner et al.~\cite{STRAW,GEISER}. 

\subsubsection{Hardware trigger rates.}
For triggering, an event must produce a voltage magnitude above 2.8$\sigma$ 
on 5 of 12 of the local antennas within a node. This threshold 
and coincidence level is chosen both to reduce the accidental
rates and to ensure a robust signal for reconstruction of the
event. The accidental rates are addressed in the following
discussion, and although we have not yet simulated the reconstruction
process in detail, the five 2.8$\sigma$ signals yield a joint 
probability corresponding to $5.7\sigma$, and even assuming one 
of the five is excluded for any reason, the remaining four signals have a 
joint $5\sigma$ probability.

The propagation
time across the 9m high node requires that the time window for such a 
coincidence be about 80~ns, corresponding to about 16 band-limited
temporal modes of the antenna input noise voltages ($\tau = 1/\Delta f$).
For these conditions, random coincidence rates per node per
coincidence window
are given by the cumulative binomial probability density function:
\begin{equation}
P(p,k,N_{\tau}) = \sum_k^{N_{\tau}} \frac{N_{\tau}!}{k!(N_{\tau}-k)!} ~ p^k (1-p)^{N_{\tau}-k}
\label{binCDF}
\end{equation}
where  $p$ the individual antenna probability of exceeding
a given voltage threshold in the window, $k$ the number of antennas
required above threshold, and $N_{\tau} = \frac{T}{\tau}~N_{ant}$ is the 
product of the number of antennas $N_{ant}$ 
and the number of independent temporal modes
of length $\tau$ within a coincidence window.
Since antenna noise voltages obey a nearly Gaussian voltage
distribution~\cite{Kra88}
\begin{equation}
p ~=~\frac{2}{\sqrt{2\pi}} \int_{2.8}^{\infty} e^{-x^2/2} dx~~=0.00511~.
\end{equation}
For bandwidths
of $\Delta f \simeq 200$~MHz, $\tau = 1/\Delta f\simeq 5$~ns, 
and thus the number of possible ``cells'' that can combine to form
a node trigger is $N_{\tau} = 16 \times 12 = 192$. The resulting 
probability of a single node trigger per 80~ns
is $P_{node} = 3.27 \times 10^{-3}$, yielding a node random
trigger rate of about $R_{node}=65$~kHz. 

Since this rate is too high for
sustained transient digitizer throughput, an additional 
digital signal processor is used to reduce the
rate by first-order causality constraints. 
To enforce causality for the local antenna node triggers, we require
that the hits along a node evolve along a proper light cone.
Consider a matrix of 12 antennas by the 16 independent 5~ns 
temporal modes in the 80~ns coincidence window, giving 
a 192-cell array of antenna hits vs. time. Considering
only plane waves for the moment, causality 
requires that hits evolve along the node on 
linear trajectories which must begin or end with the upper or
lower antenna. There are $\sim 16$ upgoing and $\sim 16$ downgoing
such trajectories, each of which has $P(p=0.0511,k=5,N_{\tau}=12)~=~2.67 \times 10^{-9}$
from equation~\ref{binCDF}. Any of the 32 causal trajectories can
begin in each 5~ns temporal mode, so the random rate for
such node triggers is
\begin{displaymath}
R_{node}^{~C} ~\simeq~  2.67\times 10^{-9} \frac{32}{5 \times 10^{-9}} \simeq  20~{\rm Hz}
\end{displaymath}
which provides ample margin on the expected $\sim 1$~kHz throughput of
the transient digitizer, and allows for some relaxation of the 
trigger requirements to account for the additional possible patterns when
wavefront curvature due to nearby showers is included.


\begin{figure} 
\epsfxsize=3.3in
\epsfbox{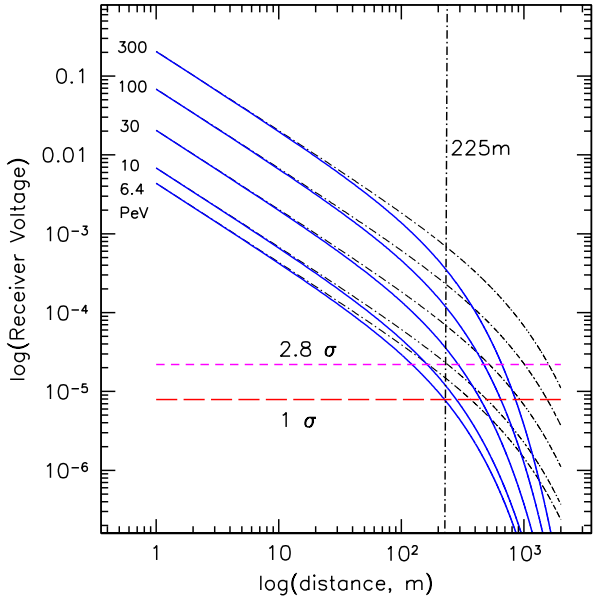}
\vspace{10pt}
\caption{Detection threshold estimates for a Saltdome Shower
Array (SalSA). Solid \& dashdot curves show received voltages as a function of 
distance for a 250~m (solid) and 900~m (dashdot) attenuation lengths, 
for the energies marked at left.
The $1,2.8\sigma$ noise levels shown are for a 450~K system temperature. The
225~m line is the antenna node spacing value used for the Monte Carlo.}
\label{SalsaThresh}
\end{figure}

Once these local node events are transmitted to the surface, global triggers
based on array-processing of event clusters are determined. An example
of how this may be implemented is as follows. For each array node
that produces a trigger, we require
a minimum of 4 additional local node triggers among 
all of its nearest neighbors including all diagonals. For each of the 
8 array corner nodes
there are 7 neighbors, for each of the 120 edge nodes there are 11, 
for each of the 600 face nodes there are 17, 
and for the remaining 1000 interior nodes there are 26. 
The light-crossing time in salt 
along the diagonal of an interior sub-cube of 27 nodes (consisting of
a center node and its 26 neighbors) is $3.23~\mu$s, or about 40 of the
node trigger windows $T_{node}=80$~ns. For a simple random coincidence, there are
$(27~{\rm nodes} \times 40~{\rm trigger~windows})~=~ 
1080$ cells in the trigger matrix. In this case, we use equation~\ref{binCDF} with
\begin{displaymath}
p= P_{node} = T_{node}~R_{node}^{~C} = 80~{\rm ns}~ \times 20~{\rm Hz~} = 1.6 \times 10^{-6}
\end{displaymath}
along with k=5, and N=1080. 

The resulting cluster trigger probability for the $3.23 \mu$s window
is $P_{cluster} = 1.27 \times 10^{-16}$, giving a net
cluster trigger rate of $R_{cluster}=3.9 \times 10^{-11}$~Hz. For the same 5-node
criterion the corresponding probabilities and rates for the faces, edges, 
and corners are far lower. Thus the potential contamination of true
events by random coincidences per year for the entire array is bounded by
\begin{displaymath}
R_{array} = P_{cluster} \times N_{interior} \times 365\times 86400~=~1.23~{\rm Hz}.
\end{displaymath}
indicating that, even prior to global event fitting constraints, the
contamination from randoms can be made negligible. In practice, 
either local thresholds or cluster trigger
criteria can be relaxed in hardware to ensure a manageable random ``heartbeat''
rate of order several Hz, and further constraints on cluster causality
can be enforced in software. This will be done to ensure adequate
real-time monitor of array instrument health.

\subsubsection{Thermal noise backgrounds.}

From above, causality requirements alone
are adequate to eliminate random triggers from the
neutrino candidate event sample. The additional information from
polarization, amplitude gradient, and
antenna waveform phase
at each node will only further establish this conclusion. The strength
of these constraints arises primarily from the fact that each
node is not just a point detector, but provides significant
directional information on its own. The important conclusion here
is that the sensitivity of our implementation is limited only
by our conservative event reconstruction requirements 
(which determine the 5 antenna + 5 node
coincidence levels), rather than random thermal noise
backgrounds.

We stress that thermal noise backgrounds in radio \v Cerenkov detectors,
while in some ways analogous to the through-going 
cosmic-ray muon backgrounds seen
in lower-energy optical \v Cerenkov detectors such as AMANDA, are
in fact quite distinct in practice, since they do not represent an
irreducible background to the physics events, and cannot reproduce
the characteristics of neutrino events.
Atmospheric muon backgrounds in lower energy detectors {\rm are} irreducible
over some range of solid angle near or above the horizon. A more 
relevant comparison with thermal noise backgrounds can be made with
downgoing muon events mis-reconstructed as upcoming neutrino-induced events
in low-energy detectors. In this
case, the frequency of mis-reconstructed events for the lower energy
neutrino telescopes can be made negligible
by choosing higher software thresholds, and the same is certainly
true for radio detectors as well with regard to thermal noise backgrounds.

\subsubsection{Simulation results.}

\begin{figure*} 
\includegraphics[width=6in]{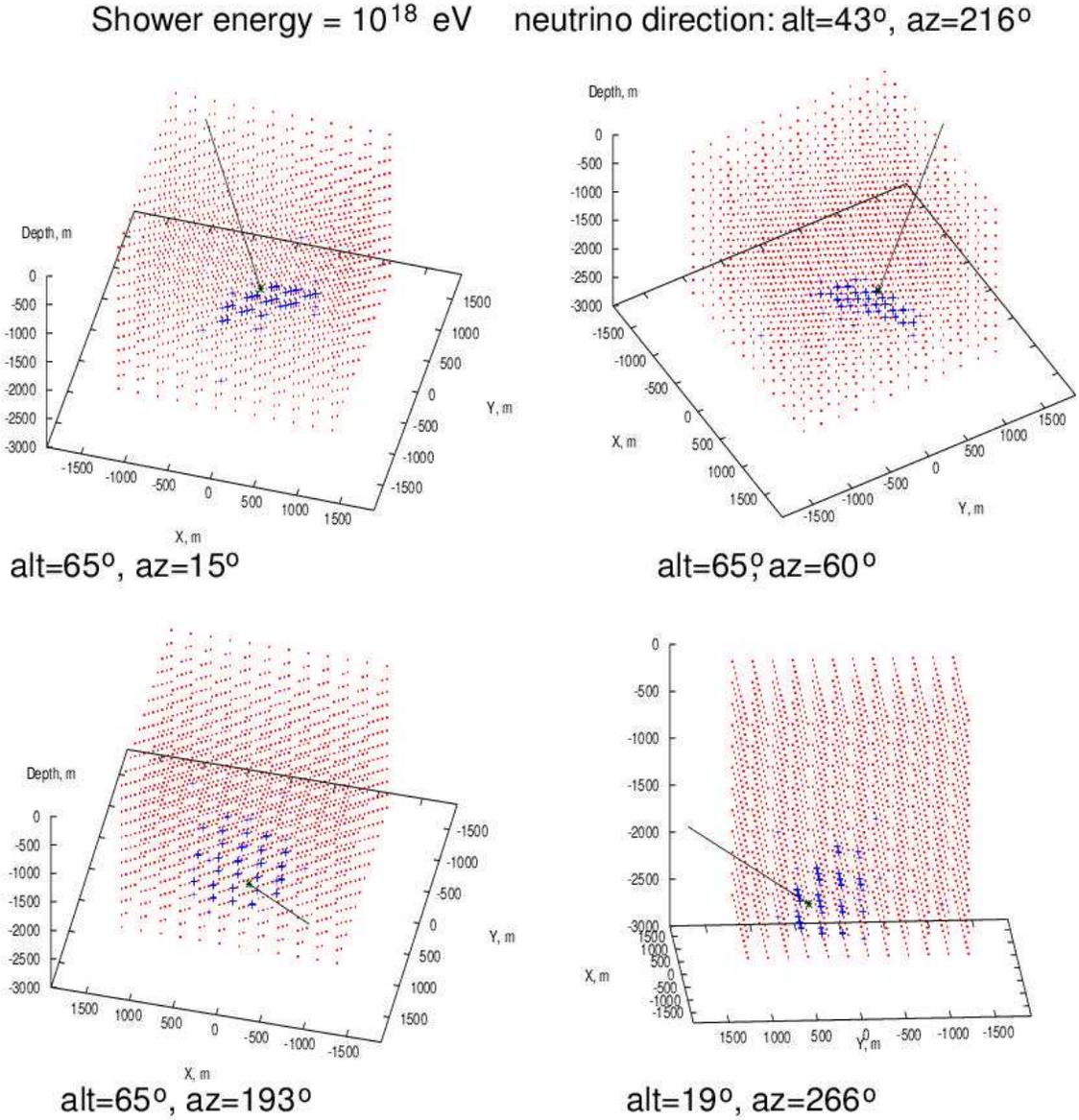}
\vspace{10pt}
\caption{Four views of the same $10^{18}$~eV hadronic shower, with
directions in local altitude and azimuth (counter-clockwise from east) shown.
The dots mark antenna nodes, and crosses mark triggered nodes, with the
incoming neutrino track shown up to its termination at the shower vertex.}
\label{Event1e18}
\end{figure*}

Figure~\ref{SalsaThresh} shows the dependence of the received voltage,
based on the parametrization of reference~\cite{Alv98,Alv97}, scaled for
salt, as a function of distance and shower energy. The curves are labeled by
shower energy in PeV, and both the $1\sigma$ and $2.8\sigma$ levels for
rms voltage above the baseline noise level are shown. The solid
and dash-dot curves  are for $L_{\alpha}=250,~900$~m respectively.

It is evident that 
the 225~m spacing we have employed in our Monte Carlo here will give a
shower energy threshold in the neighborhood of 30~PeV for our baseline
assumption of $L_{\alpha}=250$~m, and in practice we
find that other loss factors and the neutrino interaction Bjorken y-factor 
(eg., inelasticity) contribute to push the optimal 
sensitivity to neutrino energies of order
100~PeV. Figure~\ref{SalsaThresh} indicates that, although it may be possible to operate
a SalSA at a low enough energy threshold to detect events at the
6.4 PeV Glashow Resonance for $W$-vector boson production 
via $\nu_e + e^- \rightarrow W^-$, the node spacing required is
below 100~m, implying an order of magnitude more array elements. Since,
for a broad-spectrum neutrino source, the integral over the
resonance is relatively small in terms of the number of events~\cite{Gan00}
compared to the continuum,
the justification for the increased array density is not clear.
In addition, the GZK neutrino spectrum peaks above $\sim 100$~PeV.
However, if $L_{\alpha} \gg 250$~m as indicated by recent results,
the energy threshold will be considerably lower, of order
10~PeV at a spacing of 225~m, and only a modest increase in array
density required to achieve good sensitivity at the Glashow resonance.

\begin{figure*} 
\epsfxsize=6.8in
\includegraphics[width=6in]{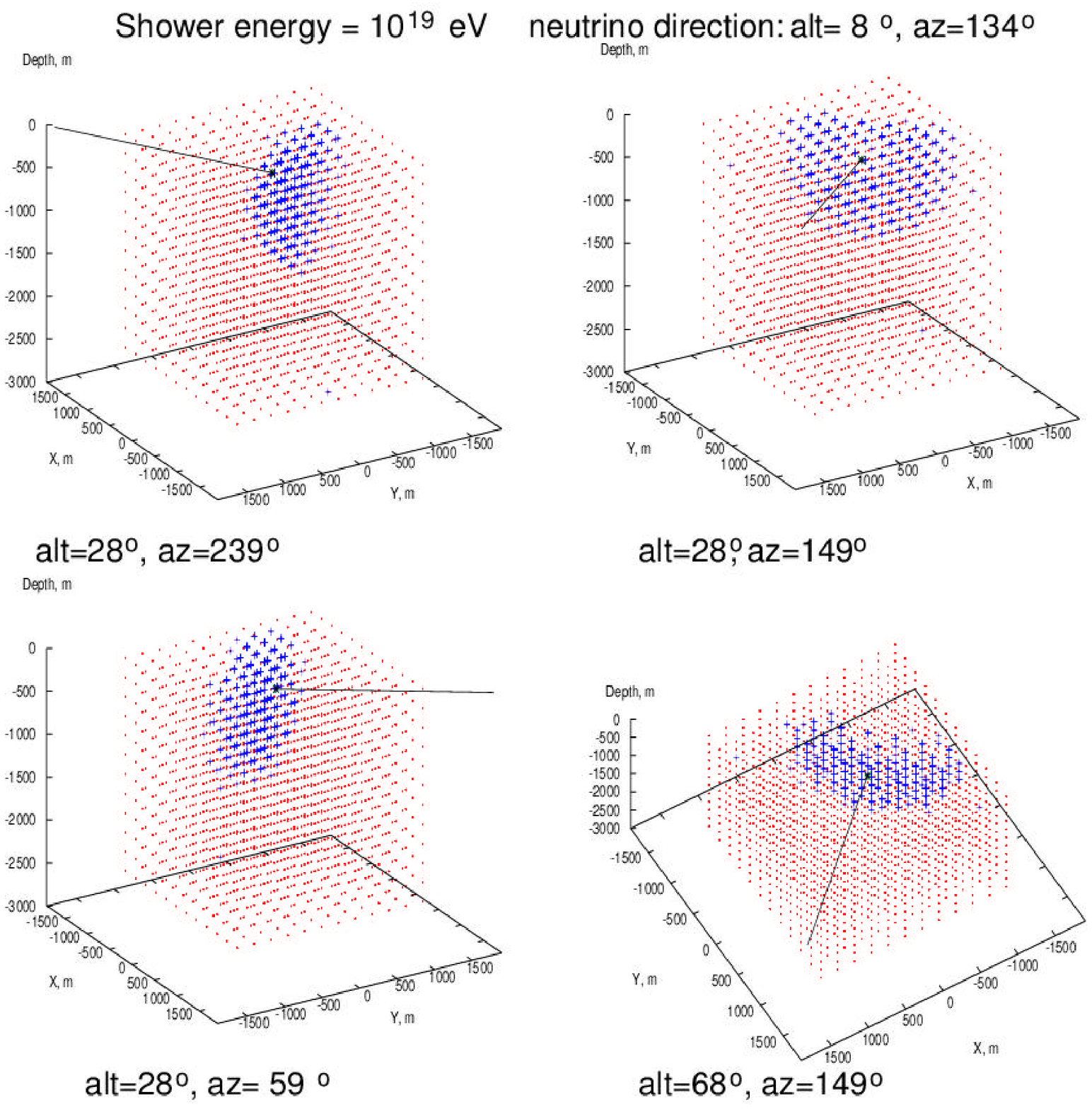}
\vspace{10pt}
\caption{Similar to previous, with shower energy $E_{sh}=10^{19}$~eV.}
\label{Event1e19}
\end{figure*}

\begin{figure} 
\epsfxsize=3.5in
\epsfbox{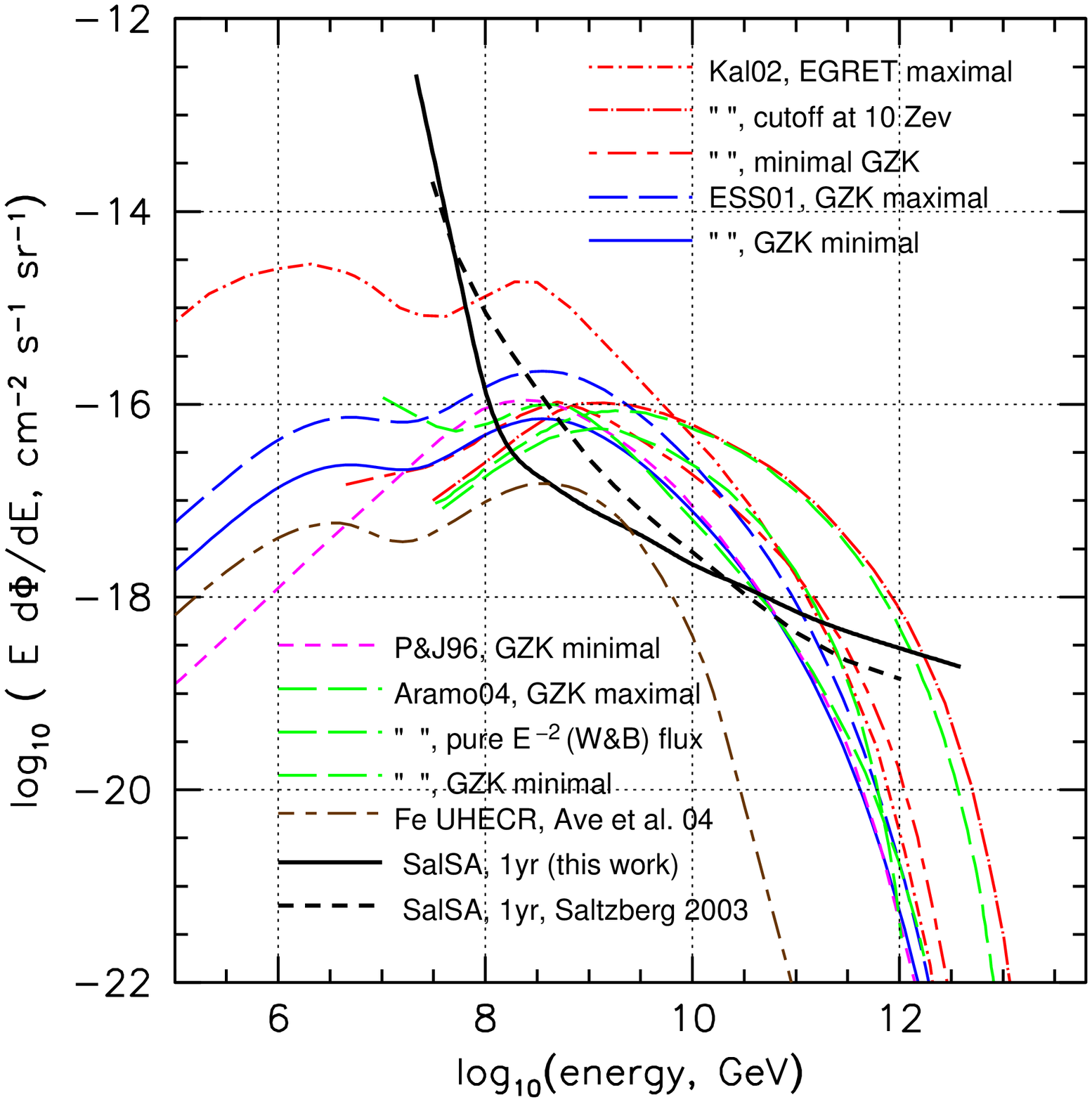}
\vspace{10pt}
\caption{Broad-spectrum flux sensitivity for a SalSA for one year of
exposure, estimated from a Monte Carlo simulation. A complete set of
GZK models is shown, as well as an earlier estimate of SalSA 
sensitivity~\cite{Saltzberg03}. From the highest to lowest the
legend indicates fluxes from Kalashev et al. (2002)~\cite{Kal02}
for the maximal level allowed by EGRET limits, an intermediate
model with a 10~ZeV cutoff energy for the UHECR spectrum, and a
minimal GZK model. The two models marked ``ESS01'' are
from Engel et al. 2001~\cite{Engel01} for one minimal model
and one with strong source evolution. The chosen P\&J96 model~\cite{PJ96}
is for a minimal GZK flux. Three different models are shown from
Aramo et al. 2004~\cite{Aramo04}, a maximal model subject only
to observational constraints, an intermediate model which follows
a pure $E^{-2}$ spectrum as described by ref.~\cite{WB}, and another
minimal GZK model. Finally, a model based on the unlikely
assumption that the UHECR composition
is pure iron is shown~\cite{Olinto04}.}
\label{gzksalsa}
\end{figure}

The volume enclosed by the simulated array
is approximately 15.6 km$^3$ of salt, or about 34 km$^3$ water equivalent mass. 
An additional 11.4(63) km$^3$ (24.9(137) km$^3$ w.e.) is contained in the boundary 
of the array within one attenuation length (for $L_{\alpha}=250(900)$~m) 
of the outer wall, where we have indicated the perimeter values for the longer
attenuation length in parentheses. These perimeter events will in
general also yield events which are well measured when they illuminate a face 
of the array. The effective fiducial volume thus approaches 60(170) km$^3$ w.e.
for an array that easily fits into the upper $\sim 3$~km of a typical salt dome.
The solid angle acceptance of the array includes the entire upper 2$\pi$ hemisphere,
and extends about $10-15^{\circ}$ below the horizon, depending on
neutrino energy. This implies of order 400(1190) km$^3$~sr w.e. acceptance,
with a threshold of order $10^{17}$~eV or less. Clearly an accurate measure 
of the attenuation lengths is still an important factor in the design,
but our goal here is to show that even conservative values for $L_{\alpha}$
give compelling results.

The simulation itself is a straightforward Monte Carlo integral of the
acceptance, using an isotropic neutrino flux, and integrating through a
shell-density model for the earth to determine the angular distribution of
neutrinos that can interact within the detector. We
use a 1:1:1 $e:\mu:\tau$ mix of neutrino flavors, and include neutral and
charged current events. For the radio emission, the Zas-Alvarez-Mu\~niz
parametrization~\cite{Alv97} is used. Full vector polarization is implemented. 
An noted above, we have not yet studied the event reconstruction 
precision, but this is not essential to determining the sensitivity,
once the appropriate instrumental threshold is established.

In Figures~\ref{Event1e18},~\ref{Event1e19} we show 4 views each of 
two different events with shower energies of $10^{18}$ and $10^{19}$~eV. 
Here the dots locate antenna nodes, and those that produce triggers
are marked with an 'x.' The incoming neutrino track is also indicated,
with the shower occurring at its termination within the detector.
The high refractive index of salt leads to a large \v Cerenkov angle of
about $66^{\circ}$ and accounts for the apparent ``flatness'' of
the cone of triggered regions around the event vertices. These events
are not selected to be typical but rather to illustrate the detector
geometry. Events of lower energy,
while able to be reconstructed via both event geometry and polarization
information, are not as easily discerned by eye.

\subsubsection{Estimates of GZK neutrino event rates.}
The Monte Carlo is evaluated at discrete energies over the PeV to ZeV
energy range, assuming a monoenergetic neutrino flux at each energy. The
values of this function at discrete energies trace a smooth curve which
represents a model-independent estimate of the sensitivity for fluxes
which are smooth compared to the energy resolution of the instrument,
which is of order $\Delta E / E \simeq 1$, limited primarily by the
uncertainty in inelasticity, which has a mean of order 0.23 at
these energies and a standard deviation about equal to its mean.
The resulting curve gives the acceptance, in km$^2$~sr, which can be multiplied 
by the livetime to determine the exposure in km$^2$~sr~s. The inverse
of the exposure gives the model-independent flux sensitivity using
the procedure outlined in ref.~\cite{Anch02}.

The result of this estimate for the flux sensitivity for 1 year of
exposure is shown in Fig.~\ref{gzksalsa}, plotted along with various
estimates of GZK neutrino fluxes~\cite{Engel01,Kal02,PJ96,Aramo0}
for a complete range of possible
parameters. A model which just equaled the sensitivity curve 
for one decade of energy would give about 3 events.
Also plotted is an earlier SalSA 
sensitivity estimate~\cite{Saltzberg03} with a simpler detector model
and more conservative trigger but otherwise similar methodology.
For GZK neutrino models we have included estimates from four independent
calculations, and in several of these cases we provide both minimal and
maximal estimates where they were made.

For the three minimal
GZK models shown the expected rates are about 11-14 events per year.
If the observed rates are found to be significantly below this, 
the physics implications for
both GZK neutrinos and cosmic rays would be serious as there are
no currently self-consistent models which predict such low rates. One
class of models in which the UHECR are predominantly heavy nuclei such as
iron have been considered~\cite{Olinto04}, and such models produce a somewhat lower
GZK neutrino rate, leading to of order 5 events per year in our
simulated detector. The assumption of no light nuclei among
the UHECR is however purely {\it ad hoc} at this stage, and the data do not
support such an inference, but the importance of this study is that
it demonstrates that there are no currently known ways to evade
the production of GZK neutrinos, even for assumed cosmic ray compositions
that are artificially designed to do so.

GZK neutrino models in the mid-range that assume stronger source evolution or
a higher-energy cutoff to the cosmic ray source spectrum
yield event SalSA rates in the range of 20-65 events per year; all of
these models are still consistent with the Waxman-Bahcall
limit for optically thin sources distributed cosmologically
in a manner similar to AGN~\cite{WB}. For the highest GZK neutrino models in
which all parameters are maximized subject only to firm observational
constraints such as the EGRET gamma-ray limits, the possible
event rates are up to 120 per year.

We note that the event rates for the minimal GZK neutrino flux models, 
while still in the regime of small statistics, are likely to 
be measured with no physics background. 
The only possible background that we currently know of
comes from charm quark production in ultra-high energy cosmic rays, which
can decay to energetic secondary muons, which can then shower via either
bremsstrahlung or photonuclear processes in the detector. However, 
even the highest current estimates~\cite{charm} for this process yield
throughgoing muon rates of $\leq 0.02$ per year above $10^{17}$~eV
over a 2.4~km radius area circumscribed around the detector volume.
With regard to electromagnetic interference,
even just several meters of overburden of rock or soil above a saltdome
is enough to attenuate all anthropogenic radio or electronic
interference, so that the system sensitivity will be subject only to 
the absolute thermal noise floor. 

For a system with a design lifetime of 10 years or more, 
even minimal GZK neutrino fluxes will produce a rich data sample
of events which would allow good precision on estimates of the
GZK neutrino energy spectrum and sky distribution. In addition,
the possibility of measuring secondary showers from the energetic
leptons in charged-current events compared to neutral current events
would allow for full calorimetry of the neutral and charged-current channels,
since at EeV energies, secondary lepton are very likely to shower
at least 10\% of their energy within 1-2~km. Landau-Pomeranchuk-Migdal
effects on the development and radio emission pattern from electron-neutrino 
events~\cite{Alv01} can potentially allow flavor
identification, as can decays of tau leptons produced in tau neutrino
events~\cite{doublebang}. 

Dramatic increases in neutrino cross sections
above the standard model,
such as might be produced by the effects of microscopic black hole
production in theories with large extra dimensions~\cite{blackholes}, 
would yield much  larger event rates than described here;
several thousand events per year are then possible even for 
minimal GZK models. Events so produced would also have quite distinct
character compared to the dominant charge-current deep-inelastic
neutrino events, since the decay of the black hole by Hawking
radiation leads to a pure hadronic shower with no secondary 
high energy leptons to produce more extended showers. There
would also be no significant LPM effects in electron neutrinos, assuming the black
hole production cross section dominated over the standard
model cross sections. In fact, high cross sections combined
with standard model GZK neutrino fluxes would serve to
yield a detection with a proportionally smaller detector.

\section{Conclusions}

We conclude the following from our measurements and simulations:
\begin{enumerate}
\item The Askaryan effect behaves as theoretically expected in measurements made
in synthetic rock salt. Correlation of the strength of the radiation to the
estimated charge excess is excellent.

\item The polarization properties of the observed radio emission are
consistent with coherent \v Cerenkov radiation, and can be used to derive
geometric properties of the shower track to good precision.

\item The observed radiation is completely coherent over many decades of shower energy,
and a wide range of radio and microwave frequencies.

\item Closely related transition radiation from the Askaryan charge excess of
showers exiting dielectric or metallic media has been measured and found to
be in agreement with predictions.

\item An antenna array of order 2.5 km on a side in an underground salt dome
is well able to conclusively detect and characterize all current standard
model GZK neutrino fluxes on even a 1 year time scale; over 10 years of
operation, good statistics can be obtained on the GZK neutrino spectrum
above $10^{17}$~eV.

\end{enumerate}

\section*{Acknowledgements}
We thank the staff of the Experimental Facilities Division at SLAC for their
excellence and professionalism in support of our efforts, M. Rosen and
his team of student technicians at the Univ. of Hawaii; D. Besson, J. Learned,
A. Odian, and W. Nelson for advice and useful discussion.
This work was performed in part at the Stanford Linear Accelerator Center,
under contract with the US Dept. of Energy, and
at UCLA under DOE contract DE-FG03-91ER40662, and under DOE contract
DE-FG03-94ER40833 at the University of Hawaii. Both P. Gorham and D. Saltzberg
are grateful for the support of DOE Outstanding Junior Investigator
awards for studies in radio detection of high energy particles.

\section*{Appendix A: Details of transition radiation analysis}

The forward spectrum of transition radiation is given in~\cite{Gor00} as
\begin{equation}
{{d^2W_{TR}}\over{d\omega d\Omega}} = {{h\alpha}\over{2\pi^3}}
{{\sqrt{\epsilon_2} \sin^2\theta \cos^2\theta}\over
{1 - \beta^2 \epsilon_2 \cos^2 \theta}} |\zeta|^2
\end{equation}

where $\zeta$ is given by

\begin{equation}
\zeta = {{(\epsilon_2 - \epsilon_1) (1 - \beta^2\epsilon_2 - \beta
\sqrt{\epsilon_1 - \epsilon_2 \sin^2 \theta})}\over {(\epsilon_1 +
\sqrt{\epsilon_2} \sqrt{\epsilon_1 - \epsilon_2 \sin^2 \theta})
(1 - \beta \sqrt{\epsilon_1 - \epsilon_2 \sin^2 \theta})}}
\end{equation}

$\omega$ is the bandwidth of the antenna ($d\omega = 2\pi d\nu$), and
$\Omega$ is the solid angle subtended by the horn, given by
$\Omega = A/d^2$ where $d$ is the distance from the horn to the source and
$A$ is the area of the horn. $\theta$ is the angle between the beamline
and the line to the horn. $\epsilon_1$ and $\epsilon_2$ are the upstream
and downstream complex dielectric constants 
($\epsilon_k = \epsilon_k^{\prime} + i \epsilon_k^{\prime\prime}$), respectively. 
For these two experiments,
the upstream medium is aluminum ($\epsilon_1^{\prime} \sim 10$) and the downstream
medium is air ($\epsilon_2^{\prime} = (1.00035)^2$).

The energy is obtained by multiplying the RHS of equation 1 by the
bandwidth and solid angle. To convert to power, divide the energy by
the typical time of the pulse (about the inverse bandwidth of the antenna).
The power delivered to the load, $P_L$, is this power 
multiplied by the horn efficiency, about 0.5 ~\cite{Kra88}.

It is useful to look at the electric field $E$ as well as the power.
The Poynting flux $S$ is given by $S = P_L/A_{eff} = E^2/377$, where
$A_{eff}$ is the effective area of the antenna. Therefore $E = \sqrt{377 P_L/A_{eff}}$.
We wish to look at the peak electric field, so we multiply $E$ by a factor of
$2\sqrt{2}$, since $E_{peak} = 2\sqrt{2} E_{RMS}$ for a perfectly triangular envelope.

The finite beam size introduces coherence corrections. 
The corrected power is given by
\begin{equation}
P = N_e (1 + N_e f_L f_T \chi)P_0
\end {equation}

where $f_L$, $f_T$, and $\chi$ are the longitudinal, transverse and
angular form factors, respectively. If the distribution of the electrons
is Gaussian, then Shibata {\it{et al}}~\cite{Tak00} give the form factors as:
\begin{equation}
f_L = \exp[{-(\pi \sigma_L \cos \theta / \lambda)^2}]
\end{equation}

\begin{equation}
f_T = \exp[{-(\pi \sigma_T \sin \theta / \lambda)^2}]
\end{equation}

\begin{equation}
\chi = \left ( {{2\theta^2}\over{\pi \Omega^2}}
\int_0^{\pi/2\theta}{x[(1 - x) K(y) + (1 + x)E(y)] e^{ {-{{\theta^2}\over
{\Omega^2}}} x^2} dx} \right )^2
\end{equation}

where $\sigma_L$ and $\sigma_T$ are the longitudinal and
transverse divergence parameters, $\Omega$ is the angular divergence
parameter, and $K(y)$ and $E(y)$ are the complete elliptic 
integrals of the first and second kinds, where $y = 2\sqrt{x} / (1+x)$:

\begin{equation}
K(y) = \int_0^{\pi/2}{(1 - y^2 \sin^2 \omega)^{-1/2} d\omega}
\end{equation}

\begin{equation}
E(y) = \int_0^{\pi/2}{(1 - y^2 \sin^2 \omega)^{1/2} d\omega}
\end{equation}

In the perfectly coherent limit, $f_L = f_T = \chi = 1$ 
and since $N_e \gg 1$, $P = N_E^2P_0$.

To compare the measured power to the calculated power, we 
take the average of $V^2$ over the
measured pulse and divide by 50$\Omega$. The measured voltage must be 
corrected for the attenuation in the circuit: 
$V_{correct} = V_{meas} 10^{atten(dB)/20}$.

To get the measured electric field, take the peak electric 
field $E_{peak} = 2 V_{peak}/h_{eff}$
where $h_{eff}$ is the effective height and $V_{peak}$ is the measured peak voltage,
with a factor of 2 to account for voltage division into a matched load.

\section*{Appendix B: Details of \v Cerenkov radiation analysis}

Zas, Halzen, and Stanev~\cite{ZHS92} provided a
detailed analysis and simulations of coherent \v Cerenkov radio emission
from high energy showers in ice. A variety of studies since then have
refined these initial results and extended them to various scenarios of
neutrino flavor and energy~\cite{Alv00a,Alv00b}. For our analysis,
we use the same approach as in our previous SLAC experiment~\cite{Sal01},
and adapt the parameterizations for the field strength 
produced by a shower in ice to the following form for salt:
\begin{equation}
R | \mathbf{E} (\nu, R, \theta_{c}) | 
= A_{0}~f_{d}~\kappa~ \left ( \frac {E_{sh}} {1~{\rm TeV}} \right )  \frac{\nu}{\nu_{0}}  
\left ( \frac{1}{1+( \frac {\nu}{\nu_{1}} )^{\delta}} \right )
\end{equation}
where R is the distance from the charged-particle beam in \v Cerenkov cone direction.
$A_{0}$ and $\delta$ are empirically determined coefficients, $A_{0} = 
2.53 \times 10^{-7}$ and $\delta = 1.44$. The factor $f_{d}$ is a scaling 
factor for the difference of radiation length and density between 
salt and ice, $f_{d} = 0.52$, and $\kappa\approx 0.5$ 
is a factor accounting for antenna near-field effects, 
with an estimated systematic uncertainty 
of $\sim 30\%$. Here $E_{sh}$ is the shower energy and is given by
$E_{sh} = E_{e} N_{e} t$, where $E_{e} = 28.5$~GeV 
is the energy of the electrons in the beam, $N_{e}$ is the number of electrons 
and $t$ is the thickness of the radiator in $X_{0}$. 
The parameters $\nu_{0} = 2200$~MHz, and $\nu_{1} = 3500$~MHz  
determine the decoherence behavior, and are estimated from the parameters for
ice, scaled by the ratio of the radiation 
length in ice and salt, calculated from the empirical formula:
\begin{equation}
 X_{0} = 716.4 \frac {\sum_{i} A_{i}} {\sum_{i} (Z_{i} (Z_{i} + 1) 
 \ln \frac {287} {\sqrt{Z_{i}}})}~~~ \lbrack \frac {g} {cm^{2}} \rbrack
\end{equation}
where $Z_{i}$ is the atomic number and $A_{i}$ is the atomic weight of the elements 
which form the molecule.


\end{document}